\documentclass[aps,twocolumn,secnumarabic,balancelastpage,amsmath,amssymb,nofootinbib]{revtex4}
\usepackage{color}           
\usepackage{graphics}         
\usepackage[pdftex]{graphicx}  
\usepackage{longtable}        
\usepackage{epsfig}
\usepackage{bm}             
\usepackage{lineno}
\usepackage{thumbpdf}
\usepackage[colorlinks=true]{hyperref}  
\begin{document}
\title{Improved Thermoelectric Cooling Based on the Thomson Effect}
\author{G. Jeffrey Snyder$^{1*}$, Eric S. Toberer$^2$,  Raghav Khanna$^1$,  Wolfgang Seifert$^3$}
\date{\today}
\affiliation{$^1$Materials Science, California Institute of Technology, 1200 E. California Blvd. Pasadena, CA 91125, USA}
\affiliation{$^2$Department of Physics, Colorado School of Mines, Golden CO 80401, USA}
\affiliation{$^3$Institute of Physics, University Halle-Wittenberg, D-06099 Halle, Germany}

\begin{abstract}
 Traditional thermoelectric Peltier coolers exhibit a cooling limit which is primarily determined by the figure of merit, \textit{zT}. Rather than a fundamental thermodynamic limit, this bound can be traced to the difficulty of maintaining thermoelectric compatibility. Self-compatibility locally maximizes the cooler's coefficient of performance for a given $zT$ and can be achieved by adjusting the relative ratio of the thermoelectric transport properties that make up $zT$. In this study, we investigate the theoretical performance of thermoelectric coolers that maintain self-compatibility across the device. 
We find such a device behaves very differently from a Peltier cooler, and term self-compatible coolers ``Thomson coolers'' when the Fourier heat divergence is dominated by the Thomson, as opposed to the Joule, term.   A Thomson cooler requires an exponentially rising Seebeck coefficient with increasing temperature, while traditional Peltier coolers, such as those used commercially, have comparatively minimal change in Seebeck coefficient with temperature.  When reasonable material property bounds are placed on the thermoelectric leg, the Thomson cooler is predicted to achieve approximately twice the maximum temperature drop of a traditional Peltier cooler with equivalent figure of merit ($zT$). We anticipate the development of Thomson coolers will ultimately lead to solid state cooling to cryogenic temperatures.\\
[1ex] 
 \hfill  PACS numbers: 84.60.Rb, 05.70.Ce, 72.20.Pa, 85.80.Fi
\end{abstract}

\maketitle

\section{Introduction}

Peltier coolers are the most widely used solid state cooling devices, enabling a wide range of applications from thermal management of optoelectronics and infra-red detector arrays to polymerase chain reaction (PCR) instruments.  Thermoelectric coolers have been traditionally understood by means of the Peltier effect, which describes the reversible heat transported by an electric current.   This effect is traditionally understood in terms of absorption or release of heat at the junction of two dissimilar materials.  The conventional analysis of a Peltier cooler approximates the material properties as independent of temperature (Constant Property Model (CPM)).  
This results in a maximum cooling temperature difference $\Delta T_{max}$ for a CPM cooler, which dependent on the figure of merit $ZT$ of the device  \cite{GoldsmidBook,Heikes}.
\begin{equation}  \label{Eq_dTmax_CPM}
\Delta T_{max} = \frac{ZT_c^2}{2}
\end{equation}

For the best commercial materials this leads to a $\Delta T_{max}$ of ~65K (single stage) \cite{Marlow}, which translates to a device $ZT$ at 300K of 0.74.  In the CPM the device $ZT$ is equal to the material $zT$. Material $zT$ depends on the Seebeck coefficient ($\alpha$), temperature ($T$), electrical resistivity ($\rho$), and thermal conductivity ($\kappa$), $zT = \frac{\alpha^2 T}{\rho \kappa}$.
In the CPM, the only way to increase $\Delta T_{max}$ for a single stage is to increase $zT$, leading to the focus of much thermoelectric research on improving $zT$.
It is well known that even further cooling to lower temperatures can be achieved using multi-stage Peltier coolers \cite{GoldsmidBook,Heikes}.  
In principle, each stage can produce additional cooling to lower temperatures, regardless of the $zT$ of the thermoelectric material in the stage.  In practice, the thermal losses and complications of fabrication limit the performance of such devices. 
The 6-stage cooler of Marlow achieves a $\Delta T_{max}$ of 133\,K; this doubling of $\Delta T_{max}$ compared to a single stage cooler is achieved despite using materials with similar $zT$ \cite{Marlow}.  Alternatively, such $\Delta T_{max}$ with a single-stage CPM cooler would require $ZT$ to be 2.5.  
 
The transport properties across a single thermoelectric leg can be manipulated to improve cooling performance, although it has been less effective in reducing $\Delta T_{max}$ than a multi-stage approach. 
One common strategy is to engineer a change in extrinsic dopant concentration across a thermoelectric element which can significantly alter $\alpha$, $\rho$ and even $ \kappa$. For example, this has been demonstrated for thermoelectric generators in n-type PbTe doped with I \cite{Gelbstein2005}. Similar efforts have been done with cooling materials, as has been reviewed in ref \cite{KuznetsovCRC}.
The simplest explanation for an improvement is an in increase in the local $zT$ at some temperatures by spatially adjusting the dopant composition within a material \cite{MahanJAP91}. 
 
Early theoretical work by Sherman et al for TEC found that different $\Delta T_{max} $ could be predicted from materials have the same or similar average $zT$ but different temperature dependence of the individual properties $\alpha, \rho, \kappa$ \cite{Sherman1960}.  This demonstrated that optimizing cooler performance is significantly more complex than simply maximizing $zT$.  More recently, M\"uller et al. \cite{mueller2003,muellerpssa2006,mueller06} and Bian et al. \cite{bian2006,bian2007} used different  numerical approaches to predict substantial gains in cooling to $\Delta T_{max}$ from functionally grading where an average $zT$ remains constant in an effort to determine the best approach to functionally grading.  

Different material classes optimized for different temperatures can also be segmented together to improve performance of thermoelectric generators but the current must also be matched \cite{Moizhes1962}. The analysis of segmentation strikingly demonstrates that increasing the average $zT$ does not always lead to an increase in overall thermoelectric efficiency and so an understanding of the thermoelectric compatibility factor is needed to explain device performance \cite{snyder2004}.

This paper derives the cooling limit for a single stage, fully optimized (self-compatible) TEC that functions as
an infinitely staged cooler.   The Fourier heat divergence in such an optimized cooler is found to be dominated by the Thomson effect rather than the Joule heating as in traditional Peltier coolers. This new opportunity presents a new challenge for material optimization based on compatibility factor rather than only $zT$.

\begin{figure} 
\epsfig{file=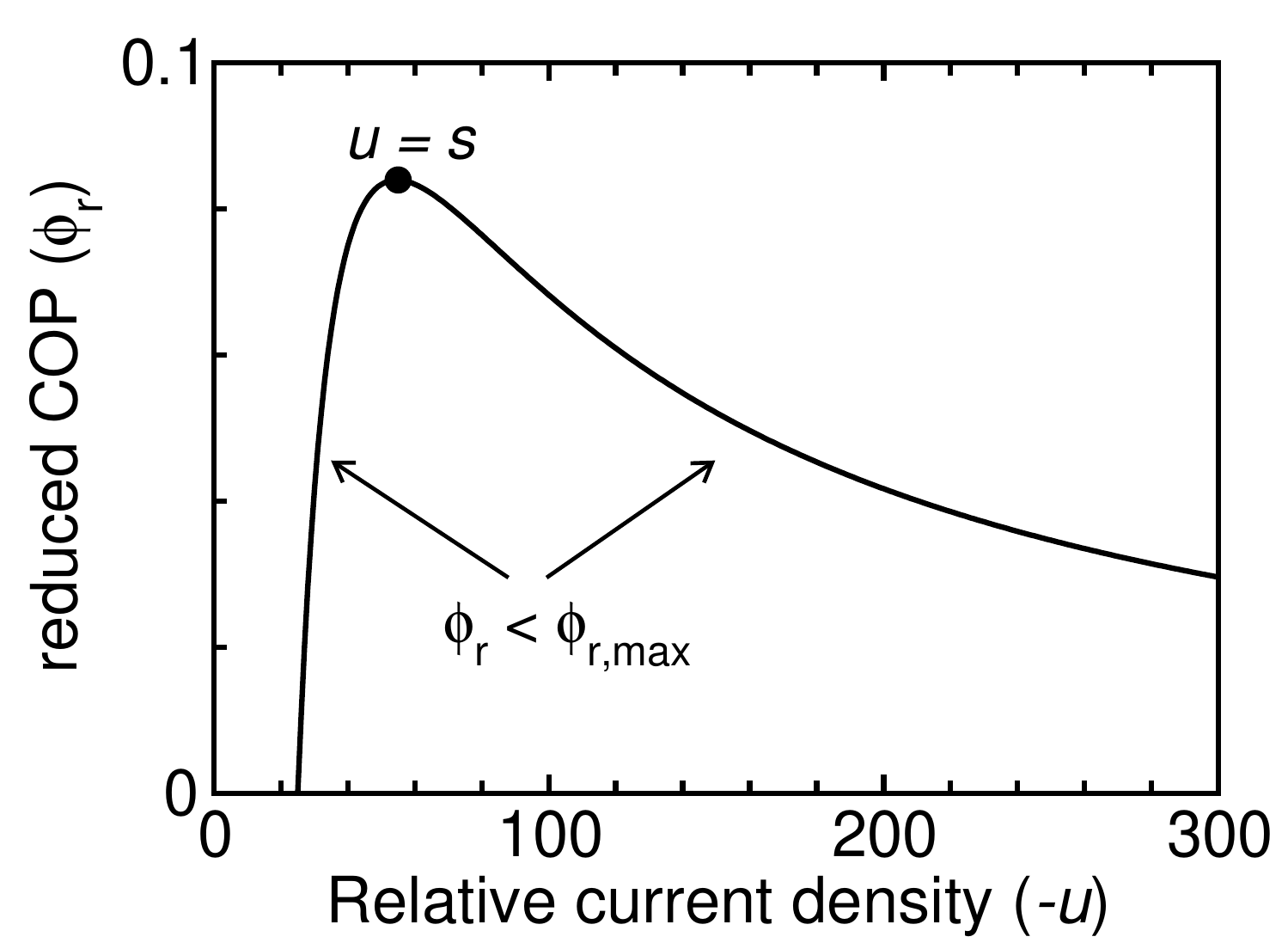, width=6cm}\\
\caption{The local reduced coefficient of performance $\phi_r$ is optimized at a specific reduced current density, termed $s$.  If $u \neq s$, the $\phi_r$ is less than that predicted by the material $zT$. Here, z = 0.002 K$^{-1}$, $\alpha = 200 \mu V K^{-1}$. }
\label{fig1}
\end{figure}

\begin{figure} 
\epsfig{file=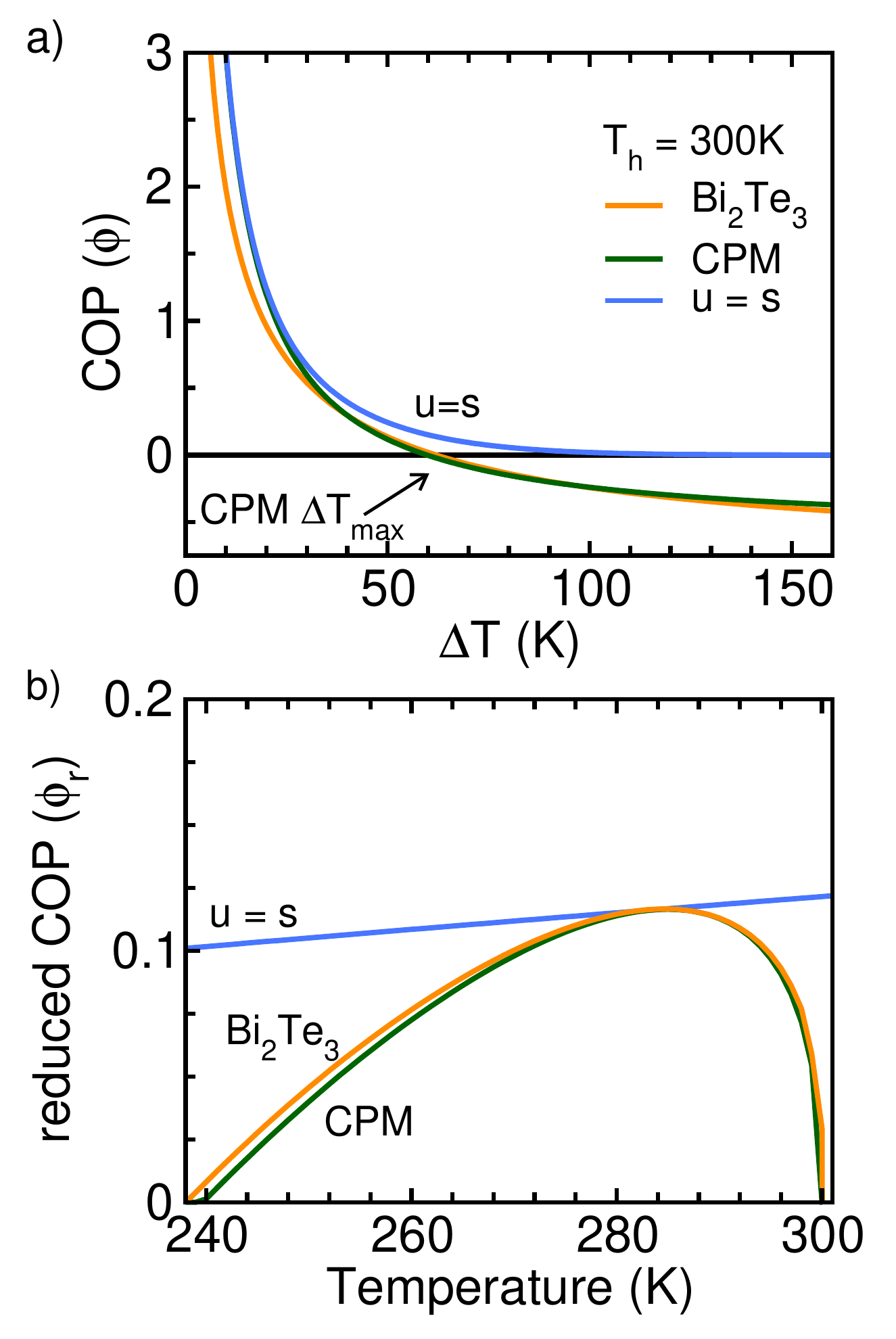, width=5.5cm}
\caption{a) The CPM Peltier cooler and $u=s$ Thomson cooler are compared using the same constant $z=0.002$\,K$^{-1}$.  The overall device $\phi$ of a CPM cooler crosses zero at a finite temperature, indicating  $\Delta T_{max}$ is reached, while $\phi$ remains positive for all temperatures for the $u=s$ cooler. b) The local performance of a CPM cooler ($\phi_r$) is significantly compromised at both the hot and cold ends.  In contrast, $\phi_{r,max}$ is achieved at all temperatures when $u=s$. In both panels, the performance calculated for an actual  Bi$_2$Te$_3$ Peltier cooler leg is similar to the CPM.}
\label{fig2}
\end{figure}


\section{Theory}

Coolers are characterized by the coefficient of performance ($\phi = Q_c / P $), 
which relates the rate of heat extraction at the cold end $Q_c$ to the power consumption $P$ in the device \cite{SnyderCRC}.   
For simplicity, but without loss of generality, a single thermoelectric element can be considered rather than a complete device.  A TEC leg can be treated as an infinite series of infinitesimal coolers, each of which is operating locally with some COP.
 Scaling this COP to the local Carnot COP ($T/dT$) yields the local reduced coefficient of performance $\phi_r$. \cite{seifertJMR2011}. 
  This relationship between local performance across the leg and global COP, $\phi$, given in Eq.\,\ref{Eq_COP} is derived in the Appendix based on Ref.\cite{Sherman1960}\cite{ZenerEgli1960}. 
 While TECs are traditionally analysed using a global approach, we have previously shown the utility of a local approach \cite{snyder03,snyder2004,SnyderCRC}.   
This local approach leads to a consideration of material `compatibility', as discussed below. 
\begin{equation}\frac{1}{\phi}= \exp{\left(\int^{T_h}_{T_c}\frac{1}{T}\frac{1}{\phi_r(T)}dT\right)}-1  
\label{Eq_COP}
\end{equation}

The compatibility approach to optimizing thermoelectric cooling arises naturally from an analysis of the thermal and electric transport equations.   
This method has been described in detail for thermoelectric generators \cite{SnyderCRC} and coolers\cite{seifert06a} and are reproduced here for TEC. The method has been experimentally verified \cite{crane2009} and shown to reproduce results using a more traditional finite element results but with less computational complexity. This method has been incorporated into several engineering models such as those used by NASA for Radioisotope Thermoelectric Generators \cite {NASATechBriefNPO-45252} and Amerigon/BSST for automotive applications \cite{crane2011,crane2009}. 
Consider an infinitesimal section of thermoelectric leg in a temperature gradient and an electric field.  
The temperature gradient will induce a Fourier heat flux ($\textbf{q}_\kappa = -\kappa \nabla T$) across this segment. The divergence of this heat (Eq.\,\ref{heateq}) is equal to the source terms: 
irreversible Joule heating ($\rho j^2$) and the reversible Thomson heat ($T \frac{d\alpha}{dT} j \nabla T$),  both of which depend on the electric current density ($j$).  
From these two effects, the governing equation for heat flow in vector notation is

\begin{equation}  
\nabla  \cdot \textbf{q}_\kappa = \nabla \cdot (-\kappa \nabla T)=\rho j^2 - \tau \,\mathbf{j} \cdot \nabla T
 \label{heateq}
\end{equation}
with  Joule heat per volume $\rho j^2$, Thomson coefficient $\tau=T 
\frac{d\alpha}{dT}$ and Thomson heat per volume $\tau \,\mathbf{j} \cdot \nabla T$.
The Peltier, Seebeck and Thomson effect are all manifestations of the same thermoelectric property characterized by $\alpha$. The Thomson coefficient ($\tau=T\frac{d\alpha}{dT}$) describes the Thomson heat absorbed or released when current flows in the direction of a temperature gradient. 

Restricting the problem to one spatial dimension, Eq. \eqref{heateq} is typically examined assuming the heat flux and electric current are parallel \cite{SnyderCRC}.
In the typical CPM model used to analyze Peltier coolers, the Thomson effect is zero 
because $\alpha$ is constant along the leg ($\frac{d \alpha}{dT} = 0$).

The exact performance of a thermoelectric leg with $\alpha(T)$, $\rho(T)$, and $\kappa(T)$ possessing arbitrary 
temperature dependence can be straightforwardly 
computed using the reduced variables: relative current 
density ($u$) and  thermoelectric potential ($\Phi$)  \cite{snyder03}.  The relative current density $u$, given in Eq.\,\ref{Eq_u}, is primarily determined by the electrical current density $j$, which is adjusted to achieve maximum global COP. 
 The thermoelectric 
potential $\Phi$ is a state function which simplifies Eq.\,\ref{Eq_COP} to  Eq.\,\ref{phifromPhi} \cite{SnyderCRC}.
\begin{equation}  \label{Eq_u}
u = \frac{-j^2}{\kappa \nabla T \cdot \textbf{j}} 
\end{equation}

\begin{equation}\label{Eq_TEpotential}
\Phi = \alpha T + 1/u
\end{equation}

\begin{equation}  \label{phifromPhi}
\phi = \frac{\Phi(T_c)}{\Phi(T_h)-\Phi(T_c)} 
\end{equation}

Changing variables to $T$ via the monotonic function $x(T)$, Eq.\,\ref{heateq} simplifies to the differential equation in $u(T)$.
\begin{equation}  \label{deqrelcurrent}
  \frac{du}{dT} = u^2 \left( T \frac{d \alpha}{dT} + \frac{\alpha^2}{z}  u \right)
\end{equation}
Using this formalism, the reduced coefficient of performance ($\phi_r$) can be simply defined for any point in the cooler  (Eq.\,\ref{Eq_eta_r}).   Fig.\,\ref{fig1} shows this relationship between $u$ and $\phi_r$. From Eq.\,\ref{Eq_COP}, it can be shown that $\phi$  is largest when $\phi_r$ is maximized for every infinitesimal segment along the cooler. Hence, global maximization can be traced back to local optimization \cite{seifert2010}.  

\begin{equation} \label{Eq_eta_r}
\phi_r = \frac{ u\, \frac{\alpha}{z} + \frac{1}{z\,T}}{u 
\frac{\alpha}{z}~(1 - u \,\frac{\alpha}{z})}
        = \frac{u \alpha + \frac{1}{T}}{u(\alpha - u \rho \kappa)}
\end{equation}

The optimum $u$ which maximizes $\phi_r$ ($\frac{d\phi_r}{du}=0$) can be expressed solely in terms of local material properties (Eq.\,\ref{Eq_sc}). This optimum value of $u$  is defined as the  thermoelectric compatibility factor $s_c$ for coolers.
\begin{equation} \label{Eq_sc} 
s_c = \frac{-\sqrt{1+zT}-1}{\alpha T} 
\end{equation}
As this paper strictly focuses on coolers, we will refer to $s_c$ as simply $s$.  

The maximum local $\phi_r$, denoted $\phi_{r,max}$, occurs when $u = s$. The expression for $\phi_{r,max}$  (Eq.\,\ref{Eq_phirmax}) is an explicit function of the material $zT$ and is independent of the individual properties $\alpha$, $\rho$, $\kappa$.   This maximum allowable local efficiency provides a natural justification for the definition of $zT$ as the material's figure of merit.  
 
\begin{equation} \label{Eq_phirmax} 
\phi_{r,max}=\frac{\sqrt{1+zT}-1}{\sqrt{1+zT}+1} 
\end{equation}
One thus wishes to construct devices where, locally, each segment has ``$u=s$" and thus  $\phi_{r,max}$ is obtained.  Globally, maximum $\phi$ is found when the entire cooler satisfies $u=s$.


\section{Cooling Performance}
To compare the cooling performance of traditional Peltier coolers and $u = s$ coolers, we consider coolers with equivalent $z$.  
Traditional Peltier coolers have typically been analyzed with the constant property model (CPM), yielding a constant $z$ (where $zT$ is linearly increasing with temperature).  We will show that constant $z$, but allowing $\alpha$, $\kappa$, $\rho$  to vary with $T$, can lead to substantial improvement in cooling.  At the limit of this variation, we will assume the properties can be varied to satisfy $u = s$.   

\paragraph*{Performance of a CPM cooler} 
CPM coolers have been extensively studied, typically using a global approach to the transport behavior.  
The $\phi$ for a CPM cooler (operated at optimum $j$) is given by Eq.\,\ref{COP_CPM} \cite{heikes1961}.  Figure \ref{fig2}a  shows the $\phi$ of a CPM cooler decreases with increasing $\Delta T$.   With increasing cooling, this $\phi$ decreases and reaches zero at $\Delta T_{max}$ (Eq.\,\ref{Eq_dTmax_CPM}). 
  
\begin{equation}
\phi^{CPM} = \left( \frac{T_c}{\Delta T} \right) \left( \frac{ \sqrt{1+zT_{avg}}-\frac{T_h}{T_c}}{ \sqrt{1+zT_{avg}} +1}  \right)
\label{COP_CPM}
\end{equation}

To understand what is limiting the CPM cooler at $\Delta T_{max}$, we derive the local reduced coefficient of performance $\phi_r^{CPM}(T)$.  To obtain $\phi_r^{CPM}$ we need $u$ as a function of $T$.  The solution to differential equation\,\ref{deqrelcurrent} for CPM is
\begin{equation} 
\frac{1}{u(T)^2}=\frac{1}{u_h^2}+\frac{2\alpha^2}{z}(T_h-T) 
\label{CPM_u}
\end{equation}
where the value of $u$ at $T=T_h$ ($u_h$) serves as an initial condition.  This expression allows $u(T)$ to be determined for any CPM cooler, regardless of temperature drop ($\Delta T \leq \Delta T_{max}$) and applied current density (\textbf{j}).  The global maximum COP ($\phi$) is obtained when the optimum $u_h$ from Eq.\,\ref{maxu_h} is employed.

  \begin{equation}
 \frac{1}{ u_h}=\frac{-\alpha}{z}\frac{z T_c^2-2(T_h-T_c)}{T_h+T_c\sqrt{z(\frac{T_h+T_c}{2})+1}} 
\label{maxu_h}
\end{equation}

\begin{figure} 
\epsfig{file=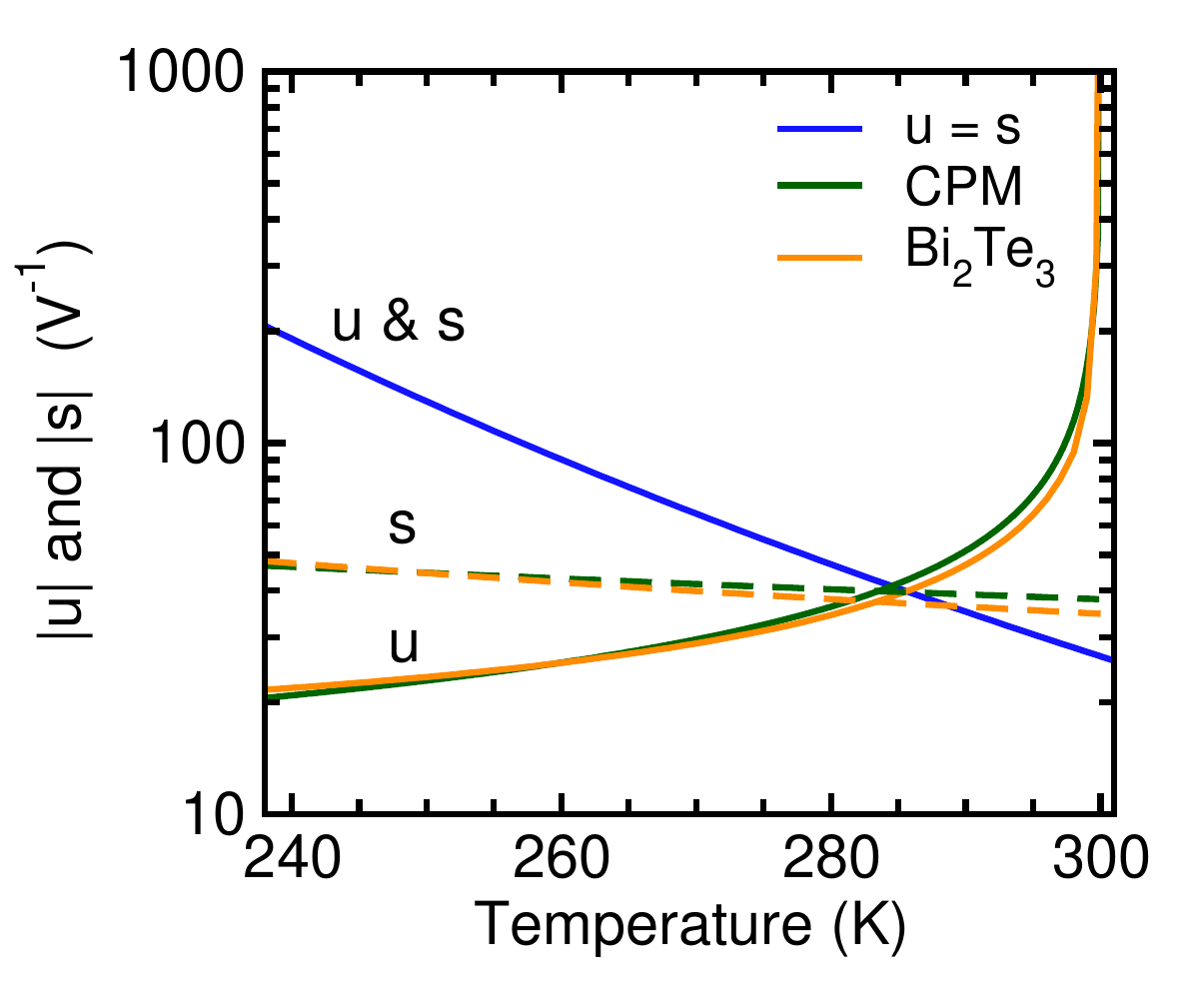, width=6cm}\\
\caption{Both the CPM and Bi$_2$Te$_3$ coolers have $u = s$ at one point along the leg. By definition, the $u=s$ is self-compatibile along the entire leg.  The different slope signs of $u$ for CPM and $u = s$ reveals that these coolers are fundamentally distinct.  The curves were generated for an optimized cooler at $\Delta T_{max}$ with $z = 0.002$ K$^{-1}$,   $T_h = 300$ K.  }
\label{fig3}
\end{figure}

Consideration of Eq.\,\ref{maxu_h} reveals that the maximum $T_c$ is obtained when $1/u_h$ approaches zero.  Figure\,\ref{fig3} shows $|u|$ becoming infinite at $T_h$ for the CPM cooler.  In this limit, Eq.\,\ref{maxu_h}  can be simplified to give Eq.\,\ref{Eq_dTmax_CPM} with  $Z=z$.   Thus, a local approach to transport yields the classic CPM limit typically obtained through an evaluation of global transport behavior.

Combining Eq.\,\ref{Eq_eta_r}, \ref{CPM_u}, and \ref{maxu_h}  results in $\phi_r(T)$  at $\Delta T_{max}$ for the CPM Peltier cooler (Eq.\,\ref{CPM_phidtmax}).  
This expression reveals $\phi_r$ drops to zero  at both ends of the CPM cooler leg, as shown in Figure\,\ref{fig2}b.  This  prohibits additional cooling and sets $\Delta T_{max}$.

\begin{equation} \label{CPM_phidtmax}
\phi^{CPM}_{r,\Delta T_{max}}=\frac{\sqrt{2z(T_h-T)}-2\frac{T_h-T}{T}}{1+\sqrt{2z(T_h-T)}}
\end{equation}

  To achieve cryogenic cooling ($T_c \rightarrow 0$) within the CPM, $zT$ must approach infinity (Eq.\,\ref{Eq_dTmax_CPM}).  
  For example, cooling with a single-stage CPM cooler to $10\,K$ would require $zT$ to be over $1000$ if the hot side is $300\,K$.  When $\phi$ is negative, the net effect of the thermoelectric device is to supply heat, rather than remove heat, from the cold side. For negative $\phi$ values for the CPM cooler to be obtained requires certain parts of the cooler to locally possess $\phi_r<0$. Such a result may be surprising at first as this $\phi_r<0$ region is made from  material possessing positive $zT$. This seems particularly odd when compared to the behavior of staged generators, discussed above.  Clearly, single- and multi-staged CPM legs exhibit fundamentally different behavior, despite being composed of exactly the same material.  Such behavior can be rationalized using the thermoelectric compatibility concept.
   
Figure\,\ref{fig3} shows that the compatibility condition ($u=s$) is maintained at only one point in the CPM cooler. Consequently,  CPM coolers operate inefficiently ($u \neq s$) at both the hot and cold ends. This is demonstrated in  Figure\,\ref{fig2}b, where $\phi_{r} < \phi_{r,max}$ for all but one point.    Once $\phi_r$ goes below zero at low temperature, the thermoelectric device is no longer cooling the cold end and $\Delta T_{max}$ is reached (Figure\,\ref{fig2}a).

While real coolers do not possess temperature-independent properties, the qualitative results for CPM translate well to traditional Peltier coolers due to their weak material gradients. Considering a Bi$_2$Te$_3$ leg with temperature-dependent properties described in Ref.  \cite{seifert2002}, we find $u$ and $s$ to be quite close to a $z$-matched CPM cooler (Figure\,\ref{fig3}). Like the CPM cooler, $u=s$ at only one temperature along the leg.  This leads to similar $\phi_{r}(T)$ for the Bi$_2$Te$_3$ and CPM coolers, shown in  Figure\,\ref{fig2}b.

Within the CPM, large $zT$  results in a high upper limit to $\phi_r$ but does not ensure this $\phi_{r,max}$ is achieved. Generaly, commercial cooling materials such as Bi$_2$Te$_3$ and any material that can be described by the CPM model will be operating   significantly below the $\phi_r$ predicted by the $zT$ they possess (Eq. \ref{Eq_eta_r}, \ref{Eq_phirmax}).  \\

\paragraph*{Performance of a $u=s$ cooler} 
We now consider an idealized cooler which maintains $u=s$ across the entire leg. $\phi_r$ for this cooler is simply given by  Eq.\,\ref{Eq_phirmax}. This $\phi_r$ is found to be positive for all $T$, as $z$ is always a positive real number.  Globally, this translates to the analytic maximum for $\phi$ for a cooler where $z$ is defined and limited.  

To facilitate comparison with CPM, we consider a constant $z$ model where the individual properties are adjusted to maintain $u=s$.  The constant $z$ approach yields vanishing $zT$ at low $T$, consistent with real materials. 
Evaluating $\phi$ (Eq.\,\ref{Eq_COP}) for a $u=s$ cooler and the assumption of constant $z$, one obtains Eq.\,\ref{COP_const_z},  where $M_i = \sqrt{1+zT_i}$ with $T_i = T_h$, $T_c$. 
\begin{equation}  \label{COP_const_z}
\frac{1}{\phi^{u=s}}=\left(\frac{M_h-1}{M_c-1}\right)^2 \exp{ \left( \frac{2(M_h-M_c)}{(M_h-1)(M_c-1)} \right)}-1
\end{equation}
Inspection of Eq.\,\ref{COP_const_z}, where $M_h > M_c > 1$, 
reveals that $\phi$ is always greater than zero for a $u=s$ cooler.

The difference between CPM and $u = s$  coolers can be visualized in  Fig.\,\ref{fig2}a, with the $\phi$ of the Thomson cooler asymptotically approaching zero with increasing $\Delta T$.  Figure\,\ref{fig2}b shows that $\phi_r$ for a self-compatible cooler with constant $z$ remains finite and positive throughout the device.  In contrast, the CPM cooler is operating inefficiently at both the hot and cold ends, limiting its temperature range. 

In principle, if $u = s$ can be maintained, the idealized $u = s$ cooler can achieve an arbitrarily low cold side temperature as long as the all of the materials have a finite $zT$.  However, the material requirements to maintain $u = s$ become exceedingly difficult to achieve as the cooling temperature is reduced and  the ultimate cooling will be finite, yielding $T_c> 0$.\\


\section{Material Requirements}

A CPM cooler has a fixed $z$ and performance which is independent of the ratio of individual properties as long as they are constant with respect to temperature (Eq.\,\ref{Eq_dTmax_CPM}).  
In contrast, a $u = s$ cooler requires dramatic changes in properties with temperature to maintain self-compatibility.  Within the constraint of constant $z$, consideration of Eq.\,\ref{Eq_sc} suggests that the Seebeck coefficient  must be varied across the device to maintain $u = s$.   Additionally, as ${\alpha(T)} =\sqrt{z \rho(T) \kappa(T)}$ within a constant $z$ model,  the product  $\rho(T) \kappa(T)$ must also vary across the device.

The Seebeck coefficient  profile $\alpha (T)$ for a $u = s$ cooler with constant $z$ can be solved analytically. Combining Eq.\,\ref{deqrelcurrent} and $u=s$ yields the simple differential equation of $\alpha(T)$:  
\begin{equation}
   \frac{d}{dT} \left( \frac{\alpha T}{1+\sqrt{1 + z\,T}} \right)
 = T\, \frac{d \alpha}{dT} - \frac{\alpha}{z} ~\frac{1+\sqrt{1 + z\,T}}{T}~.
\end{equation}
Solving this equation yields 
\begin{equation}  \label{alphasoln}
 \alpha(T)=\alpha_0 \frac{\sqrt{1+zT}-1}{\sqrt{1+zT}}\exp{ \left( \frac{-2}{\sqrt{1+zT}-1}\right) } ~.
\end{equation}
 
With this expression for $\alpha(T)$, it is possible to evaluate $s(T)$ with Eq.\,\ref{Eq_sc}. 
 Figure\,\ref{fig3} shows the variation in $s$ required for a $u = s$ cooler with constant $z$. The self-compatible cooler modeled in Figure\,\ref{fig3} has $z= 0.002$; 60\,K of cooling results in a change in $s$ of one order of magnitude.  

     The approximation for small $zT$  yield a simple expression for $\alpha(T)$, given by Eq.\,\ref{alpha_approx}. 
\begin{equation}  \label{alpha_approx}
  \frac{d}{dT} \left( \ln \alpha(T) \right) = \frac{4}{zT^2} ~\longrightarrow ~ 
\alpha(T)  \propto \exp{ \left( \frac{-4}{zT} \right)} 
 \end{equation}
This reveals that $\alpha$ should be very large at the hot end and must decrease to a low value at the cold end.  
This exponentially varying $\alpha(T)$ required to maintain $u=s$ for constant $z$ is anticipated to be the limiting factor in real coolers and place bounds on the maximum cooling obtainable. We consider the realistic range of $\alpha$ below.

Large values of $\alpha$ are found in lightly doped semiconductors and insulators with large band gaps ($E_g$) that effectively have only one carrier type, thereby preventing compensated thermopower from two oppositely charged conducting species. Using the relationship between peak $\alpha$ and $E_g$ of \textit{Goldsmid} (Eq.\,\ref{GoldsmidSharp}) allows an estimate for the highest $\alpha (T_h)$ we might expect at the hot end, $\alpha_h$ \cite{thermalbandgap}.  Good thermoelectric materials with band gap of 1\,eV are common while 3\,eV should be feasible.  For a cooler with an ambient hot side temperature, this would suggest $\alpha_{h}$ should be $\sim$1-5\,mV/K.  Maintaining $zT$ at such large $\alpha$ will require materials with both extremely high electronic mobility and low lattice thermal conductivity.  
\begin{equation}  \label{GoldsmidSharp}
\alpha_{h}= E_g / (2 eT_{h}) ~
\end{equation}

A lower bound to $\alpha_c$ also arises from the interconnected nature of the transport properties. We require $zT$ to be finite; thus the electrical conductivity $\sigma$ must be large as  $\alpha_c$ tends to zero.  In this limit, the electronic component of the thermal conductivity ($\kappa_E$) is much larger than the lattice ($\kappa_L$) contribution and $\kappa \sim \kappa_E$.  To satisfy the Wiedemann-Franz law ($\kappa_E =  L \sigma T$ where $L = \frac{\pi^2}{3}  \frac{k^2}{e^2}$ is the Lorenz factor in the free electron limit), $\alpha_c$ has a lower bound given by Eq.\,\ref{alphac}. For example, a $z = \frac{1}{300}$ \,K$^{-1}$ and $T_c = 175$\,K results in a lower bound to $\alpha_c$ of $119\,\mu$V/K.  

\begin{equation}  \label{alphac}
\alpha_c^2  = L z T_c = \frac{\pi^2}{3}\frac{k_B^2}{e^2} z T_c ~
\end{equation}

The maximum cooling temperature $T_c$ can be solved as a function of $z$, $E_g$ and $T_h$ from equations  Eq.\,\ref{alphasoln}, Eq.\,\ref{GoldsmidSharp} and Eq.\,\ref{alphac}.  For small $z$ the approximate solution 
\begin{equation}  \label{ThomsonApprox}
 \Delta T \approx \frac{z}{8} \, T_h^2 ~\ln{ \left( \frac{E_g^2}{\frac{4}{3} \pi^2 k_B^2 \, z \, T_h^3} \right) } ~
\end{equation}
gives an indication of the important parameters but quickly becomes inaccurate for $zT$ above 0.1.

\begin{figure}[!ht]
\epsfig{file=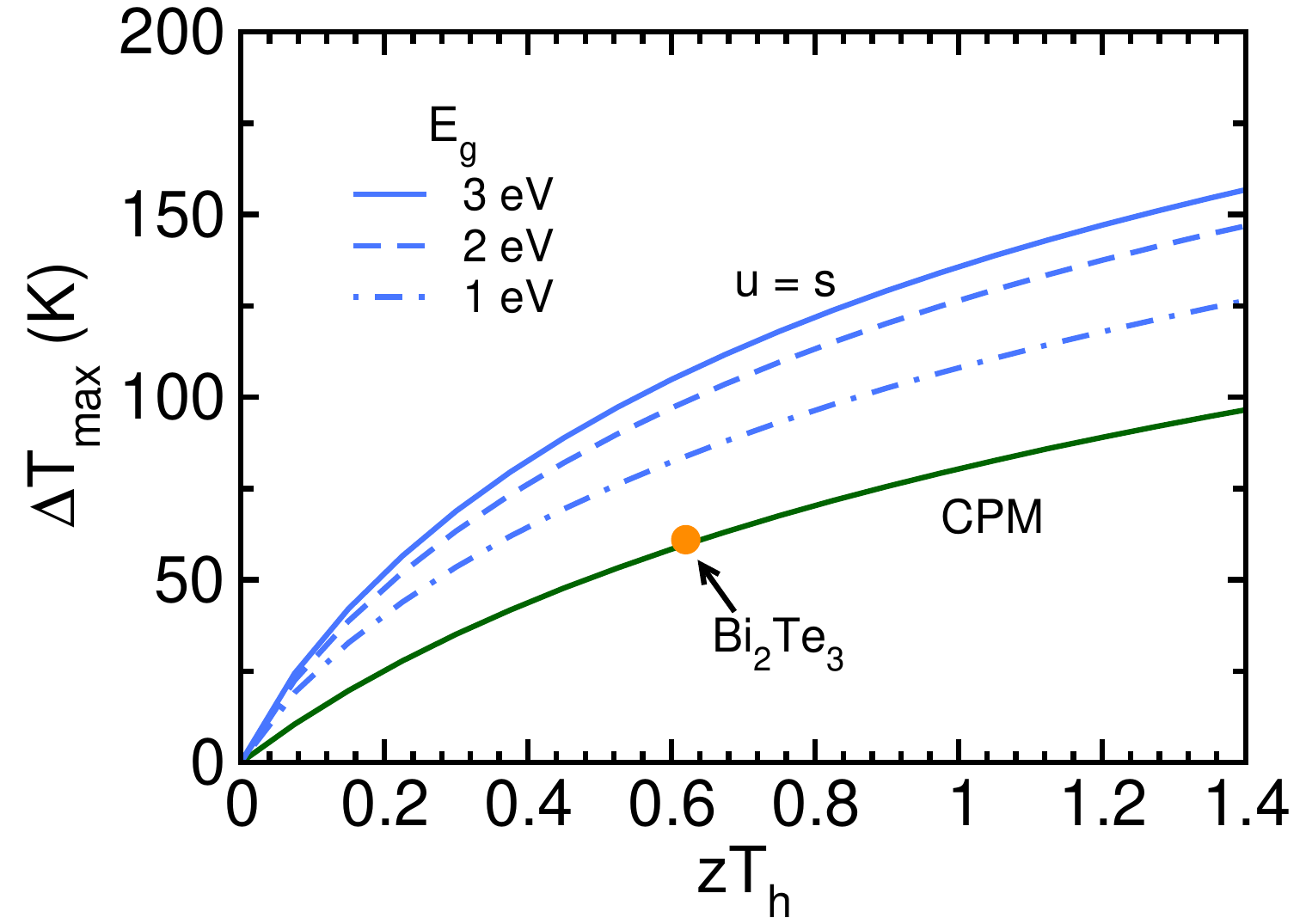, width=6.5cm}
\caption{The maximum temperature drop $\Delta T_{max}$ of a $u=s$ Thomson cooler exceeds that of a Peltier cooler with the same $z$. Large band gap, $E_g$, thermoelectric materials are necessary at the hot junction improves the performance ($T_h = 300$K).}
\label{Fig_DeltaTlimit}
\end{figure}

\paragraph*{Material limits to  performance}
With these bounds on material properties, we consider the $\Delta T_{max}$ of a $u=s$ cooler.  Figure\,\ref{fig2} suggests that the $\phi$ of a  $u=s$ cooler remains positive for all temperature.  However, obtaining materials with the required properties limits $\Delta T_{max}$ to a finite value.  Fig.\,\ref{Fig_DeltaTlimit} compares the $\Delta T_{max}$  solution for $u=s$ and CPM coolers with the same $z$. Here, the maximum Seebeck coefficient is set by the band gap  ($E_g  = 1-3~eV$), per Eq.\,\ref{GoldsmidSharp}. The $u=s$ cooler provides significantly higher $\Delta T_{max}$ 
than the CPM cooler with the same $zT$,  nearly twice the $\Delta T_{max}$ for $E_g  = 3~eV$.

\paragraph*{Spatial dependence of material properties}
These analytic results are possible because the compatibility approach does not require an exact knowledge 
of the spatial profile for the material properties. Nevertheless, it is possible solve for the spatial dependence of the $u = s$ cooler, given some material constraints.  To determine $x(T)$, we integrate Eq.\,\ref{Eq_u}, recalling we have assumed constant cross-sectional area ($j(x)=const.$), obtaining Eq.\,\ref{CRC40}.  
\begin{equation}\label{CRC40}
x(T) = \frac{-1}{j}\int_T^{T_h} u \kappa dT
\end{equation} 
Thus, the natural approach to cooler design within the $u = s$ approach is to determine the temperature dependence of the material properties, and then determine the required spatial dependence from the resulting $u(T)$ and $\kappa(T)$.

\begin{figure} 
\epsfig{file=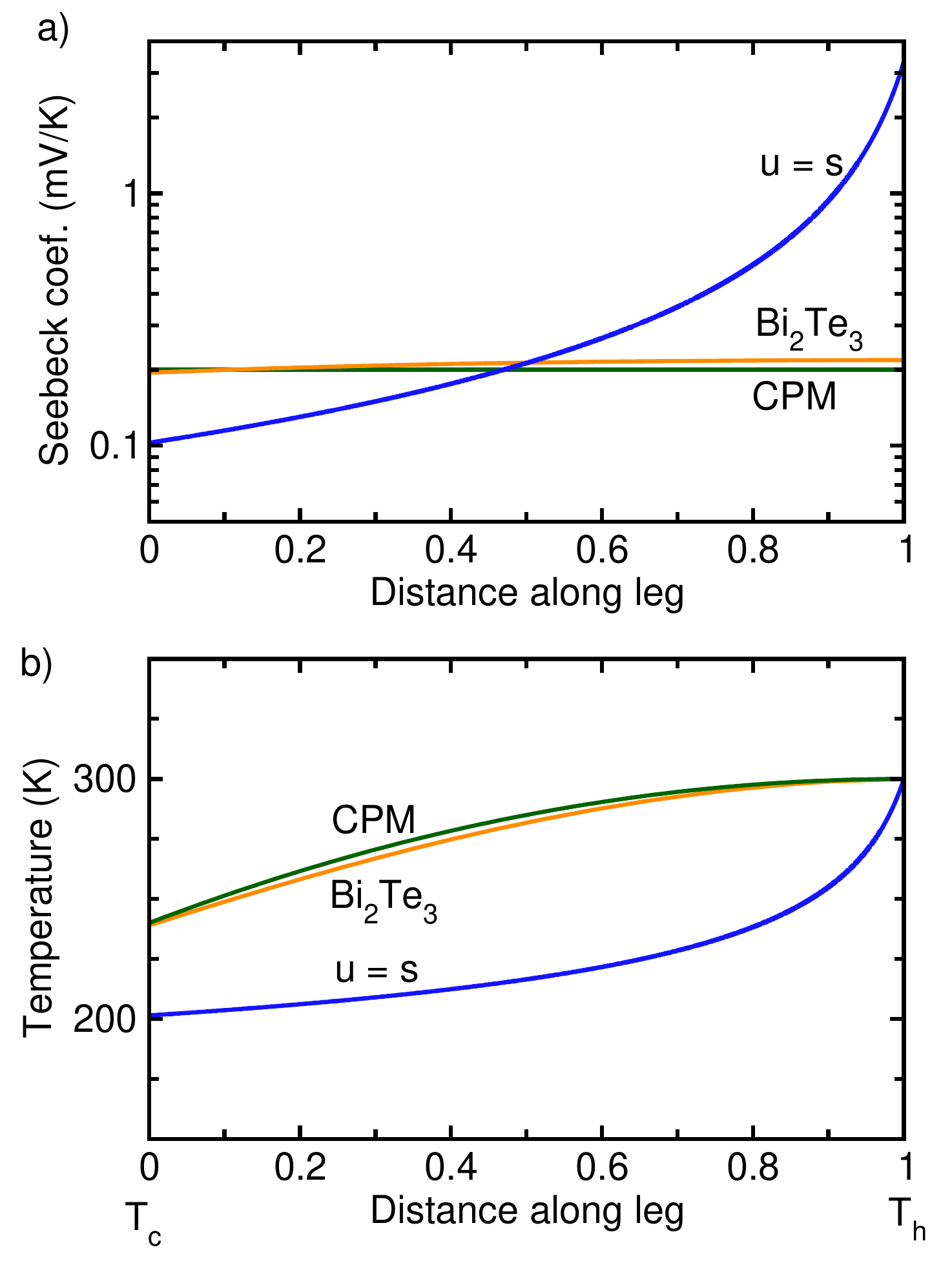, width=6cm}\\
\caption{a) The Seebeck coefficient of the $u = s$ Thomson cooler varies exponentially, while it is by definition constant for the CPM cooler. As a degenerately doped semiconductor, the Seebeck coefficient of commercial Bi$_2$Te$_3$ increases gradually with temperature.  b)  The curvature of $T(x)$  for the CPM Peltier cooler temperature profile is opposite that of the $u=s$ cooler because of the different sign of the Fourier heat divergence.  Again, similar behavior is found between the CPM and commercial Bi$_2$Te$_3$ cooler.}
\label{fig5}
\end{figure}

 Figure\,\ref{fig5}a shows an example of the Seebeck distribution $\alpha(x)$ along the leg that will provide 
the necessary $\alpha(T)$, where a constant $\kappa_L = 0.5$ W/mK is assumed. The $\alpha$ of  Figure\,\ref{fig5}a  spans the range permitted by Eq.\,\ref{GoldsmidSharp} and \ref{alphac}.

 In a real device the spatial profile of thermoelectric properties will need to be carefully engineered. 
 If this rapidly changing $\alpha(x)$ is achieved by segmenting different materials, low electrical contact resistance is required between the interfaces.  
We anticipate such control of semiconductor materials may require thin film methods on active bulk thermoelectric substrates.


\section{Cooler phase space}

The improved performance of a $u = s$ cooler is not simply an incremental improvement, but rather we find CPM and $u=s$ coolers operate in fundamentally different phase-spaces.  Here, by phase space we refer to the class of solutions defined by the sign of the Fourier heat divergence ($\nabla \cdot   \textbf{q}_\kappa$ in Equation 3).  The Fourier heat divergence in a cooler contains both the Joule $(\rho j^2)$ and Thomson $(\tau \textbf{j} \cdot \nabla T)$ terms.

 We begin by considering the Fourier heat divergence in CPM and Bi$_2$Te$_3$ coolers and then compare this behavior to $u = s$ coolers. 
 In the typical CPM model to analyze Peltier coolers, $\tau = 0$ as there is no variation in $\alpha$.  
 In a CPM cooler, $\nabla  \cdot  \textbf{q}_\kappa$  is thus greater than zero. 
 This can be seen by the downward concavity of the temperature distribution in Figure\,\ref{fig5}b.  
In a typical Peltier cooler (e.g. Bi$_2$Te$_3$), the concavity is the same as the CPM cooler and thus the divergence is likewise positive.  This is because the Thomson term is always less than the Joule term in a conventional thermoelectric cooler.

In contrast, a $u=s$ cooler changes the sign of the Fourier heat divergence such that  $\nabla  \cdot  \textbf{q}_\kappa$  is less than zero. This can be readily visualized in Figure\,\ref{fig5}b, where the concavity of the $u = s$ cooler is opposite the CPM and Bi$_2$Te$_3$ coolers.  This difference in concavity must come from the Thomson term being positive and greater than the Joule heating term.  The large magnitude of Thomson term is understandable with the exponentially rising Seebeck coefficient seen in Figure\,\ref{fig5}a.   The reversibility of the Thomson effect requires that for $\nabla  \cdot  \textbf{q}_\kappa$ to be less than zero, the hot end must have a high $|\alpha|$ relative to the cold end, and not vice versa.  This translates to a requirement for $\tau$ such that $\tau\textbf{j} \cdot \nabla T > \rho j^2$.

We can also express the Fourier heat divergence in terms of reduced variables.  
\begin{equation}
\nabla \cdot \textbf{q}_\kappa = \textbf{j} \cdot \nabla \frac{1}{u}  = \frac{- 1}{u^2} \frac{du}{dT} \textbf{j} \cdot  \nabla T
\end{equation}

Manipulation with Eq.\,\ref{Eq_u} produces a form where the sign of $u$ and directions of $j$  and $\nabla T$ are irrelevant.
\begin{equation}
\nabla \cdot \textbf{q}_\kappa =\frac{j^2}{2 \kappa u^4}\frac{d}{dT}u^2
\end{equation}
Thus the sign of $\nabla \cdot \textbf{q}_\kappa$ is determined by the sign of $\frac{d|u|}{dT}$, which is valid for both $p$ and $n$-type elements regardless of the sign of $u$.
 
 The Fourier heat divergence criterion is a convenient definition  to distinguish these two regions of thermoelectric cooling in experimental data.  The Peltier cooling region, defined by  $\nabla \cdot \textbf{q}_\kappa>0$,  is found in the phase space where $\frac{d|u|}{dT}>0$. Likewise, the Thomson cooling region defined by $\nabla \cdot \textbf{q}_\kappa<0$ is the phase space where  $\frac{d|u|}{dT}<0$.  The constant relative current  $u(T) = const.$ separates the Thomson-type and from the Peltier-type solutions to the differential equation.   In Figure\,\ref{fig3}, the CPM and $u = s$ cooler have opposite slopes, indicating these coolers exist in separate regions of the cooling phase space. This result is consistent with our discussion above concerning the concavity of $T(x)$ in Figure\,\ref{fig5}. 

For clarity, we suggest coolers which are predominately in the Thomson phase-space, $\nabla \cdot \textbf{q}_\kappa<0$  but may not have $u = s$  be referred to as ``Thomson coolers''. Similarly, ``Peltier coolers" should refer to  coolers operating in the usual  $\nabla \cdot \textbf{q}_\kappa>0$ Fourier heat divergence phase-space where Joule heating dominates.

This understanding of phase space for $u = s$ and CPM coolers enables us to hypothesize that the performance advantages of $u = s$  coolers extends to imperfect Thomson coolers.  We expect such coolers possess two primary advantages over traditional Peltier coolers.
First, for a given material $zT$, performance ($\Delta T_{max}$ and $\phi$) of the  Thomson cooler is greater (Figure\,\ref{fig2}).  The $\Delta T_{max}$  solution for the $u=s$ cooler is compared to a Peltier cooler with the same material assumption for $z$ in Fig.\,\ref{Fig_DeltaTlimit}. Here, the maximum Seebeck coefficient is set by the band gap  ($E_g  = 1-3~eV$), per Eq.\,\ref{GoldsmidSharp}. The Thomson cooler provides significantly higher $\Delta T_{max}$ 
than the Peltier cooler with the same $zT$,  nearly twice the $\Delta T_{max}$ for $E_g  = 3~eV$.
Second, in a  Thomson cooler, the temperature minimum is not limited by $zT$ explicitly  like it is in a traditional Peltier coolers.

\section{Discussion}

Efficiency improvements from staging and maintaining $u=s$ also exists for thermoelectric generators, 
but the improvement is small ($< 10\%$ compared to CPM). This is because the $u$ does not typically 
vary by more than a factor of two across the device. However, in a TEC the compatibility requirement 
is much more critical. When operating a TEC to maximum temperature difference, the temperature gradient varies 
from zero to very high values, which means $u$ will have a much broader range (Figure\,\ref{fig3}) in a TEC than in a generator. 
Thus, unless compatibility is specifically considered, the poor compatibility will greatly reduce 
the performance of the thermoelectric cooler, and this results in the $\Delta T_{max}$ limit well known for Peltier coolers.

In real materials, changing material composition also changes $zT$ so the effect of maximizing average $zT$ is difficult to decouple from the effect of compatibility.  As such, efforts which are focused on maximizing $zT$ will generally fail to create a material with $u = s$ and may only marginally increase $\Delta T_{max}$. Conversely, focusing on $u = s$ without consideration of $zT$ could rapidly lead to unrealistic materials requirements.

In this new analysis we have focused on the compatibility criterion, $u=s$, with constant $z$ (as opposed to $zT$ \cite{seifertJMR2011}) to demonstrate the differences between a Thomson and a Peltier cooler typically analyzed with the CPM model.  Generally, achieving $u=s$ in a material with finite $zT$, is more important to achieve low temperature cooling than increasing $zT$.     
 
Minor improvements in thermoelectric cooling beyond increasing average $zT$ by increasing the Thomson effect in a functionally graded material were predicted as early as 1960 \cite{Sherman1960}.  Similarly M\"uller et al. describe modest gains in cooling from functionally grading 
\cite{mueller2003,muellerpssa2006,mueller06} where material properties are allowed to vary in a constrained way such that the average $zT$ remains constant. Such additional constraints can keep the analysis within the Peltier region, preventing a full optimization to a $u = s$ solution. 
 
Bian et al, \cite{bian2006,bian2007} propose a thermoelectric cooler with significantly enhanced $\Delta T_{max}$ using a rapidly changing Seebeck coefficient in at least one region. The method of Bian et al focuses on the redistribution of the Joule heat rather than a consideration of the Thomson heat or the effect of compatibility. Nevertheless, the region of rapidly changing Seebeck coefficient would also create a significant Thomson effect and likely place that segment of the cooler in the Thomson cooler phase space while other segments would function like a CPM Peltier cooler.

In a traditional single-stage (or segmented) thermoelectric device, the current flow and the heat flow are collinear and flow through the same length and cross-sectional area of thermoelement.  This leads to the compatibility requirement between the optimal current density and optimal heat flux to achieve optimal efficiency. In a multi-stage (cascaded) device the thermal and electrical circuits become independent and so the compatibility requirement is avoided between stages.
 
A transverse Peltier cooler  also decouples the current and heat flow by having them transport in perpendicular directions. Gudkin showed theoretically that the transverse thermoelectric cooler, could function as an infinite cascade by the appropriate geometrical shaping of the thermoelement \cite{Gudkin1977}. The different directions of heat and current flow enable, in principle, an adjustment of geometry to keep both heat and electric current flow independently optimized. Cooling of 23 K using a rectangular block was increased to 35 K using a trapezoidal cross-section \cite{Gudkin1978}.

\section{Conclusion}

Here, we compare self-compatible coolers with CPM and commercial Bi$_2$Te$_3$ thermoelectric coolers.  Significant improvements in cooling efficiency and maximum cooling are achieved for equivalent $z$ when the cooler is self-compatible. Such improvement is most pronounced when the goal is to achieve maximum temperature difference, rather than high coefficient of performance at small temperature difference. Optimum material profiles are derived for self-compatible Thomson coolers and realistic material constraints are used to bound the performance.  Self-compatible coolers are found to operate in a fundamentally distinct phase space from traditional Peltier coolers.  The Fourier heat divergence of Thomson coolers is dominated by the Thomson effect, while this divergence in Peltier coolers is of the opposite sign, indicating the Joule heating is the dominant effect.  This analysis opens a new strategy for solid state cooling and creates new challenges for material optimization.

\begin{acknowledgments} 
We thank AFOSR MURI FA9550-10-1-0533 for support.  E. S. T. acknowledges support from the U.S. National Science Foundation MRSEC program - REMRSEC Center, Grant No. DMR 0820518.  
\end{acknowledgments}

\section{Appendix} 
The metric for summing the efficiency and coefficient of performance of these thermodynamic processes is not a simple summation because energy is continuously being supplied or removed so that neither the heat nor energy flow is a constant.  The derivation, attributed to Zener\cite{ZenerEgli1960}, of the coefficient of the performance Eq.\,\ref{Eq_COP} and efficiency summation metric for a continuous system in one-dimension given here is based on Zener and similar derivations given in \cite{Sherman1960}\cite{Freedman1966}\cite{Harman1968Book}\cite{seifert06a}. 

Consider $n$ heat pumps (or heat engines) connected in series such that the heat entering the $i$th pump, $Q_i$, is the same as the heat exiting the $i-1$ pump, namely $Q_{i-1}$.  Then by conservation of energy
\begin{equation}   \label{Q_n}
   Q_i=Q_{i-1}+P_i
\end{equation}
where $P_i$ is the power entering the pump $i$.
The coefficient of performance $\phi_n$ of the pump $n$ is defined by
\begin{equation}   \label{phi_n}
   \frac{1}{\phi_i} = \frac{P_i}{Q_{i-1}}
\end{equation}
This gives the recursive relation
\begin{equation}   \label{Q2_n}
   Q_i=Q_{i-1}(1+\frac{1}{\phi_i})
\end{equation}
which can be solved for $Q_n$ as
\begin{equation}   \label{Q3_n}
   Q_n=Q_{0}\prod_{i=1}^n (1+\frac{1}{\phi_i})
\end{equation}

The total power added to the system $P$ from the $n$ pumps  is
\begin{equation}   \label{powereq1}
   P = \sum_{i=1}^n P_i
\end{equation}

The coefficient of performance for the $n$ pumps, $\phi$, with heat $Q_c = Q_0$ entering from the cold side is defined by
\begin{equation}   \label{phi_totn}
   \frac{1}{\phi} = \frac{P}{Q_0}
\end{equation}

By conservation of energy (or recursive relation Eq \ref{Q_n}), the heat exiting the $n$ pumps, $Q_n$ is the heat pumped by the first pump plus the total power, Eq \ref{powereq1}.
\begin{equation}   \label{heateq1}
   Q_n = Q_0 + P
\end{equation}

Combining equations \ref{phi_totn}, \ref{heateq1}, and \ref{Q3_n}, it is straightforward to show 
\begin{equation}   \label{phi_product}
   (1+\frac{1}{\phi}) = \prod_{i=1}^n (1+\frac{1}{\phi_{i}})
   \end{equation}
This can be transformed into a summation by use of a natural logarithm
\begin{equation}   \label{phi_sum}
   ln(1+\frac{1}{\phi}) = \sum_{i=1}^n ln(1+\frac{1}{\phi_{i}})
   \end{equation}

While $1/\phi_i$ should become very small with small $\delta T_i = T_i - T_{i-1}$ the reduced coefficient of performance $\phi_{r,i}$ should remain finite 
\begin{equation}   \label{phir_n}
   \frac{1}{\phi_i} = \frac{1}{\phi_{r,i}}  \frac{\delta T_i}{T_i}
\end{equation}
Assuming a monotonic temperature distribution (for a simple TEC $T_i >T_{i-1}$ for all $i$) the sum in Eq. \ref{phi_sum} can be converted to an integral in the limit that $n\to\infty$, where $ \delta T_i = (T_h-T_c)/n$ and $ln(1+x)\to x$ for small $x$
\begin{equation}   \label{phi_integral}
   ln(1+\frac{1}{\phi}) = \int_{T_c}^{T_h}\frac{1}{T\phi_{r}(T)}dT
   \end{equation}
If the temperature distribution is not monotonic but can be divided up into monotonic segments, each of these monotonic segments can be individually transformed into integrals.
 
For a generator (as opposed to a cooler or heat pump) power is extracted rather than added at each segment in the series. Then the sign of the power in equation \ref{Q_n} is negative for a generator with the efficiency given by $\eta = -P/Q_0 = -1/\phi$ and reduced efficiency $\eta_r =  -1/\phi_r$. Thus the above method can be used to derive the analogous equation for generator efficiency: 
\begin{equation}   \label{eta_integral}
   ln(1-{\eta}) = -\int_{T_c}^{T_h}\frac{\eta_{r}}{T}dT
   \end{equation}




\bibliographystyle{naturemag}
\bibliography{toberer}

\end{document}